\documentstyle[twoside,fleqn,espcrc2]{article}

\newcommand{\PSbox}[3]{\mbox{\rule{-0.2in}{#3}\includegraphics{#1}}}%\hspace{#2}}}

\newcommand{\AmS}{{\protect\the\textfont2
  A\kern-.1667em\lower.5ex\hbox{M}\kern-.125emS}}

\title{Study of Instanton Contributions to Moments of Nucleon Spin-Dependent Structure Functions}

\author{D.Dolgov\thanks{Based on the poster presented by D.Dolgov.
                       Work supported by the U.S. Department
                       of Energy (DOE) under cooperative research agreement DE-FC02-94ER40818.}
          , R.Brower, %\address{Center for Theoretical Physics, Laboratory for Nuclear Science, \\
        J.W. Negele 
	and A.Pochinsky\address{Center for Theoretical Physics, Laboratory for Nuclear Science, \\
        Massachusetts Institute of Technology, \\
        77 Massachusetts Ave., Cambridge, MA 02139, USA}}
       
\begin{document}

\begin{abstract}
Instantons are the natural mechanism in non-perturbative QCD to remove helicity from valence quarks and
transfer it to gluons and quark-antiquark pairs. To understand the extent to which instantons explain the
so-called ``spin crisis'' in the nucleon, we calculate moments of spin-dependent structure functions in quenched
QCD and compare them with the results obtained with cooled configurations from which essentially all gluon
contributions except instantons have been removed. Preliminary results are presented.
\end{abstract}

% typeset front matter (including abstract)
\maketitle

\section{MOTIVATION}

In recent years there has been a major experimental effort to measure structure functions which characterize
the distribution of quarks in the nucleon. The challenge for theorists is now to understand the data which
already exist and to predict the structure functions which will be measured in currently 
planned experiments.
Lattice QCD provides the only known framework for non-perturbative calculation
of hadron structure, and although it cannot address the intrinsically
Minkowski structure functions themselves, through the operator product
expansion it is possible to calculate their moments. There have been several
calculations of these moments in recent years, including calculation of all
moments of the spin independent and longitudinal spin dependent structure
functions through order four \cite{QCDSFmain} and the tensor charge \cite{QCDSFtensor} by the QCDSF collaboration, as well as calculation of connected and 
disconnected contributions to the axial charge \cite{Fukugita,Liu} and to the
tensor charge \cite{Aoki}.
These results agree qualitatively with experiment and discrepancies may plausibly be attributed to
a combination of finite lattice size effects, and the omission of sea quarks and disconnected
diagrams.

However, in addition to showing that numerical solution of QCD reproduces the
experimental results, we want to understand as fully as possible the physical
origin of the observed structure.
Since instantons play a major role in the physics of light quarks \cite{NegInst,IvanInst},
we also would like to understand their role in hadron structure functions.
Especially interesting from this perspective is the spin structure of a proton.
The 't Hooft instanton interaction $\overline{u}_Ru_L\overline{d}_Rd_L\overline{s}_Rs_L$ is the
only known vertex in QCD that directly removes helicity from valence
quarks and transfers it to gluons and quark-antiquark pairs, thereby rendering
 instantons the natural mechanism to explain the so-called
``spin crisis''.

Hence it is important to identify the instanton contribution to the proton spin structure functions.
On the lattice, a practical way of extracting the instanton content is ``cooling''\cite{Berg,NegInst}, in which
one sequentially minimizes the action locally on each link
and iteratively approaches a stationary solution.

In this work we calculate the zeroth moments of the $g_1(x)$ and $h_1(x)$ structure functions
with and without cooling for $u$ and $d$ quarks in quenched lattice QCD. 
The  zeroth moment of $g_1(x)$ is proportional to the quark's spin contribution
to the total spin of a nucleon in the parton model 
and has been measured experimentally. The $h_1(x)$ structure
function characterizes the transversity of quarks in a proton\cite{Jaffe}.

\section{METHOD}

We use a lattice of size $16^3\times32$ with $\beta=6.0$ and generate quenched $SU(3)$ gauge fields using
overrelaxed heatbath Monte-Carlo with Cabibbo-Marinari decomposition.
We calculate the following matrix elements:
\begin{eqnarray}
\langle \vec{P}\vec{S}|\bar{q} \gamma^{i} i \gamma_5 q|\vec{P}\vec{S}\rangle &=& 2 S^i \Delta q \\
\langle \vec{P}\vec{S}|\bar{q} \sigma^{0i} i \gamma_5 q|\vec{P}\vec{S}\rangle &=& 2 S^i \delta q 
\end{eqnarray}
where $|\vec{P}\vec{S}\rangle$ is a ground state of a proton with momentum $\vec{P}$ and spin $\vec{S}$
($P^2=M_P^2$ and $S^2=-M_P^2$), and $q$ defines the quark flavor $u$ or $d$.

The axial charge is related to $g_1(x)$ through the operator product
expansion by
\begin{eqnarray*}
\lefteqn{2\int^1_0 dx\, g_1(x,Q^2) =}\\
& &\sum_{q=u,d} e^{(q)}_{1,0}(\mu^2/Q^2, g(\mu))\, Z_A((a\mu)^2, g(a))\, \Delta q(a)
\end{eqnarray*}
where $e_{1,0}$ is the Wilson coefficient and $Z_A$ is the lattice 
renormalization constant for the axial charge and analogously for
the tensor charge. To compare below with experimental data
and with the calculation of ref.\cite{QCDSFmain}, we use the perturbative value 
of $Z_A(1, g=1)=0.867$ used in their work.

The nucleon matrix elements are extracted from the ratio
\begin{equation}
\label{Rt}
R(t) = \frac{\sum_{\vec{x}} \Gamma^{\alpha\alpha^\prime}\langle J_\alpha O(\vec{x},t) \bar{J}^\prime_{\alpha^\prime}\rangle}
{\Gamma^{\alpha\alpha^\prime}\langle J_\alpha \bar{J}^\prime_{\alpha^\prime}\rangle}
\end{equation}
where $O(\vec{x},t)$ is the operator of interest, either $O(\vec{x},t)=\bar{q} \gamma^{i} i \gamma_5 q(\vec{x}, t)$ or the tensor operator;
the spin vector is chosen to be $S=(0,0,0,|S|)$; the corresponding
spin projection operator is $\Gamma = (1+\gamma^5\gamma^3)/2$;
the nucleon source is
$J_\alpha(x) = \epsilon_{abc} u_\alpha^a (u^b C\gamma_5 u^c)$;
and the states are normalized by 
$\langle \vec{P}|\vec{P}^\prime \rangle = 2E V \delta_{\vec{P},\vec{P}^\prime}$

For the hadron source ($J$) we use gaussian smearing with $\sqrt{r^2}\approx 0.5 fm$ in combination with
Coulomb gauge fixing at the time slice of the source. We use a point-like sink ($\bar{J}^\prime$) and because there is no need for finite $\vec{P}$ to
avoid operator mixing, we project onto zero spatial momentum to 
minimize numerical noise.

We evaluate the numerator in (\ref{Rt})  using  a standard sequential source
%$$\gamma^5 M^{-1}(x,y) \gamma^5 = M^{-1\dagger}(y,x)$$
to create a set of backward propagators which can be contracted with
a given operator and smeared forward propagators to produce the corresponding matrix element. 

\section{CALCULATION}

For uncooled QCD we use $\kappa=0.1515$ and $\kappa=0.1550$, where
for reference we note that for $\beta=6.0$
$\kappa_c=0.1569$ \cite{SESAMspectro,QCDSFmain}. We define the lattice spacing from 
the SESAM nucleon mass chiral extrapolation yielding
$a^{-1}=1.95\,GeV$. At this value of $a$ our $\kappa$'s correspond to $m_\pi\approx 590\,MeV$ and
$m_\pi\approx 980\,MeV$ respectively.

\begin{table*}[t]
\setlength{\tabcolsep}{2pc}
\caption{Unrenormalized axial vector charge ($\Delta u$, $\Delta d$) and 
tensor charge ($\delta u$, $\delta d$) for the un-cooled case}
\label{tab:hot}
\begin{tabular*}{\textwidth}{l||l|l|l|l}
\hline
$\kappa$ & $\Delta u$  & $\Delta d$  & $\delta u$ & $\delta d$\\
\hline
0.1515   & 1.118(43)   & -0.281(19)  & 1.297(40)  & -0.302(26)\\
\hline
0.1550   & 0.967(124)  & -0.318(70)  & 1.210(225) & -0.329(58)\\
\hline
\end{tabular*}
\end{table*}

\begin{table*}[t]
\setlength{\tabcolsep}{2pc}
\caption{Unrenormalized axial vector charge ($\Delta u$, $\Delta d$) and 
tensor charge ($\delta u$, $\delta d$) for the cooled case}
\label{tab:cool}
\begin{tabular*}{\textwidth}{l||l|l|l|l}
\hline
$\kappa$          & $\Delta u$  & $\Delta d$  & $\delta u$ & $\delta d$  \\
\hline
0.1230 & 0.814(22) & -0.230(10) & 0.969(19) & -0.236(10)\\
\hline
0.1250 & 0.703(48) & -0.232(27) & 0.916(42) & -0.222(30)\\
\hline
\end{tabular*}
\end{table*}

In the cooled case we use 25 cooling steps as in ref.\cite{IvanInst}.
Standard chiral extrapolation yields the following results:

\begin{table}[h]
% space before first and after last column: 1pc
% space between columns: 2.0pc (twice the above)
\setlength{\tabcolsep}{1pc}
% -----------------------------------------------------
% adapted from TeX book, p. 241
%\newlength{\digitwidth} \settowidth{\digitwidth}{\rm 0}
%\catcode`?=\active \def?{\kern\digitwidth}
% -----------------------------------------------------
\caption{Chiral extrapolation in the cooled case}
\label{tab:coolchiral}
\begin{tabular}{l||l|l}
\hline
$\kappa$          & $am_\pi$  & $am_N$   \\
\hline
0.1230 & 0.405(12) & 0.748(24)\\
\hline
0.1250 & 0.278(17) & 0.615(31)\\
\hline\hline
$\kappa_c$=0.1269 & - & 0.497(38)\\
\hline
\end{tabular}
\end{table}

Using $am_N$ to fix the lattice spacing we obtain $a^{-1}=2.2\,GeV$.

The three point function was evaluated with the source at $t=10$
and the sink at $t=24$, and the operator was averaged over the central plateau
from $t=12$ to $t=21$. The statistical errors in $R(t)$ in Eq.(\ref{Rt}) were
determined by the jackknife method. Because the initial sample size for the uncooled
configuration was only 40, leading to some uncertainty in the
optimum size of the central plateau, to be conservative
we added a 5\% systematic error to the purely
statistical error.

\section{RESULTS}

\begin{figure}[h]
\vspace{24pt}
\PSbox{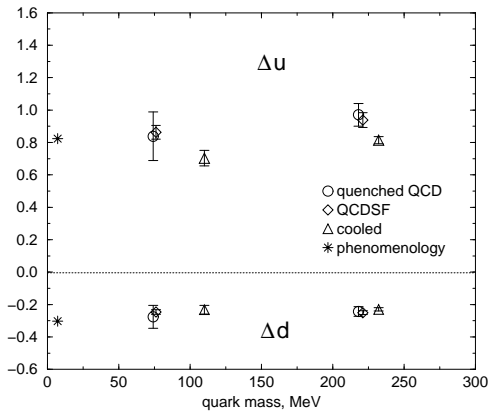 hscale=40 vscale=40}{0in}{2.0in}
%\framebox[55mm]{\rule[-21mm]{0mm}{43mm}}
\vspace{-0.5in}
\caption{Lattice results for the axial vector charges}
\label{fig:a0}
\vspace{-0.3in}
\end{figure}

Our results for the unrenormalized axial vector charge ($\Delta u$, $\Delta d$) and tensor
charge ($\delta u$, $\delta d$)
are shown in Table~1 for uncooled QCD and
in Table~2 for cooled QCD.  

In order to compare with QCDSF calculations and with phenomenological data, we consider the axial vector charge,
for which the renormalization factor $Z_A(1, g=1)=0.867$ is available, and show our
results in Figure~1. For uncooled QCD there is excellent agreement between our renormalized results
for $Z_A\Delta u$ and $Z_A\Delta d$ and the high statistics QCDSF calculation. Both lattice calculations
extrapolate well to the phenomenological data, indicating that at least for these moments,
quenched QCD is qualitatively adequate. To understand how much of the axial charge arises
from instantons alone, we also show the cooled results $\Delta u$ and $\Delta d$ in the same
figure. In principle, one should calculate renormalization factors arising from instantons.
However, although we have not yet done this, we believe that most of the renormalization
arises from fluctuations which have been removed by cooling, so that the relevant renormalization
factor is close to unity. Hence, we believe that the qualitative agreement between the cooled and uncooled
result in Figure~1 strongly suggests that instantons indeed provide the dominant contribution
to the nucleon spin.

\section{ACKNOWLEDGEMENT}

These calculations were carried out at MIT on the LNS DEC-8400 SMP,
the Pleiades Alpha 4100 Cluster, and the Sun E5000 SMP cluster and
Wildfire prototype system.

\end{document}